\documentstyle[12pt,fleqn,amssymb,epsfig,rotating]{article}

\def\lsim{\mathrel{\rlap{\lower4pt\hbox{\hskip1pt$\sim$}}
    \raise1pt\hbox{$<$}}}         %less than or approx. symbol
\def\gsim{\mathrel{\rlap{\lower4pt\hbox{\hskip1pt$\sim$}}
    \raise1pt\hbox{$>$}}}         %greater than or approx. symbol
\newlength{\dinwidth}
\newlength{\dinmargin}
\makeatletter
\def\section{\@startsection {section}{1}{\z@}{-3.5ex plus-1ex minus
    -.2ex}{2.3ex plus.2ex}{\reset@font\large\bf}}
\makeatother
\begin{document}
\begin{titlepage}
\begin{flushleft}
%
%===> Change report numbers and date
%
{\tt DESY 95-086    \hfill    ISSN 0418-9833} \\
{\tt hep-ex/yymmnn} \\
{\tt May 1995}                  \\
%{\bf\large  Draft ~ 1.05   \hspace{5mm} 03. 10. 94}  \\
\end{flushleft}
%\vspace*{4.cm}
%
% \begin{flushright}
%H1-10/94-???
%  25.\ April '95
% \end{flushright}
%\begin{flushleft}
%\vspace*{1.cm}
%Final Comments to Leif Joensson (F15JOE@DSYIBM.DESY.DE) and \\
%Joachim Stier (stier@dice2.desy.de)
%\end{flushleft}
\begin{center}
  \vspace{3cm}
  \begin{Large}
  \begin{bf}
A Direct Determination of the Gluon Density in the Proton at Low x \\
  \end{bf}
  \end{Large}
  \vspace{2cm}
  \begin{large}
  H1 Collaboration\\
  \end{large}
  \vspace{2cm}
\begin{abstract}

A leading order determination of the gluon density
in the proton has been performed
%A determination of the gluon density in the proton has been made
%from a leading order analysis
in the fractional momentum
range  $1.9 \cdot 10^{-3} < x_{g/p} < 0.18$
by measuring
multi-jet  events from boson-gluon fusion
in deep-inelastic
scattering with the H1 detector at
the electron-proton collider HERA. This direct determination of the
gluon density was performed in a kinematic region previously not
accessible.
The data show a considerable increase of the gluon density
with decreasing fractional momenta of the gluons.

\end{abstract}
\end{center}
% \end{bf}
% \end{large}
%\end{center}
%================================abstract===============================
%\vspace*{4.cm}
%{\bf Abstract:}
%\begin{quotation}
%
%\end{quotation}
%\vfill
%\cleardoublepage
\end{titlepage}

%Author list copied Monday, March 10
 S.~Aid$^{13}$,                   %HAM2-PD      8/93        Aid
 V.~Andreev$^{25}$,               %LPI-PD                   Andreev
 B.~Andrieu$^{28}$,               %ECPL-PD                  Andrieu
 R.-D.~Appuhn$^{11}$,             %DESY-PD     4/92         Appuhn
% C.~Arnault$^{27}$,              %ORSA-TP                  Arnault
 M.~Arpagaus$^{36}$,              %ZUTH-LEFT   4/95         Arpagaus
 A.~Babaev$^{24}$,                %ITEP-PD                  Babaev
 J.~B\"ahr$^{35}$,                %ZEUT-PD                  Baehr
% E.~Banas$^{6}$,                 %CRAC-TP    6/93          Banas
 J.~B\'an$^{17}$,                 %KOSI-PD                  Ban1
 Y.~Ban$^{27}$,                   %ORSa-ST                  Ban2
 P.~Baranov$^{25}$
% ...

 \vspace{2cm}

%     H1 Institutes as appearing on publications
\noindent
% ...
$ ^{11}$ DESY, Hamburg, Germany$^a$ \\
% ...
$ ^{13}$ II. Institut f\"ur Experimentalphysik, Universit\"at Hamburg,
                                     Hamburg, Germany$^ a$  \\
% ...
$ ^{17}$ Institute of Experimental Physics, Slovak Academy of
                Sciences, Ko\v{s}ice, Slovak Republic$^ f$\\
% ...
$ ^{24}$ Institute for Theoretical and Experimental Physics,
                                                 Moscow, Russia \\
 $ ^{25}$ Lebedev Physical Institute, Moscow, Russia$^ f$ \\
% ...
$ ^{27}$ LAL, Universit\'{e} de Paris-Sud, IN2P3-CNRS,
                            Orsay, France\\
 $ ^{28}$ LPNHE, Ecole Polytechnique, IN2P3-CNRS,
                             Palaiseau, France \\
% ...
$ ^{35}$ DESY, Institut f\"ur Hochenergiephysik,
                              Zeuthen, Germany$^ a$\\
 $ ^{36}$ Institut f\"ur Teilchenphysik,
          ETH, Z\"urich, Switzerland$^ i$\\
% ...
$ ^a$ Supported by the Bundesministerium f\"ur
                                  Bildung und Forschung, FRG
 under contract numbers 6AC17P, 6AC47P, 6DO57I, 6HH17P, 6HH27I, 6HD17I,
 6HD27I, 6KI17P, 6MP17I, and 6WT87P \\
% ...
$ ^f$ Supported by the Deutsche Forschungsgemeinschaft\\
% ...
$ ^i$ Supported by the Swiss National Science Foundation\\

\newpage

\section{Introduction}

Deep-inelastic lepton-nucleon scattering experiments have
played a fundamental role in reaching an understanding of
the structure of matter.  Ever since the discovery of the
proton's parton content in the late 1960s, extensive
studies have been made at accelerators providing
increasingly higher energies to obtain more detailed
knowledge of the parton properties within
nucleons. Although it was soon realized that about $50\%$
of the nucleon momentum was carried by gluons, a direct
measurement of their momentum distribution has so far
been restricted to large fractional momenta.  Instead
information about the gluons has been extracted from
measurements of the sea-quark distribution. This relies
on the assumption that sea-quark pairs are produced by
quark-antiquark production from gluons in an evolution
process.

%Instead one has extracted information on
%the gluons from measurements made on
%the sea-quark distribution, relying
%upon the assumption that sea-quark pairs
%are produced by gluon decays in
%an evolution process.
%%starting with gluon emission from the valence quarks.

\vspace*{-5mm}
\begin{figure}[hbt]
\begin{center}
\mbox{\begin{turn}{0}
\epsfig{file=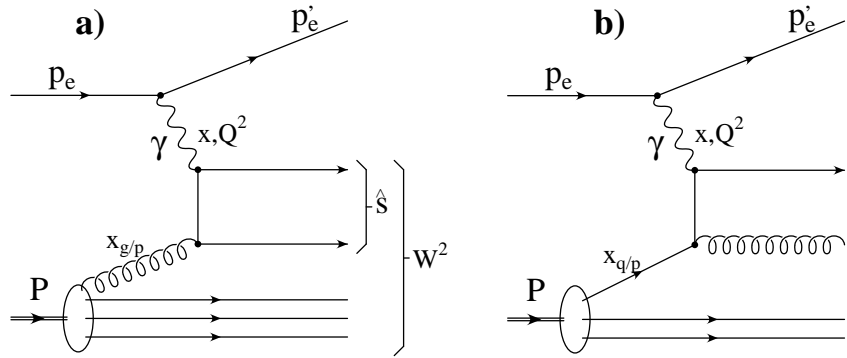,width=.7\linewidth}\end{turn}}
\caption[feynman]{{\it
Generic Feynman diagrams for a) the boson-gluon fusion process
and, b) the QCD-Compton process}}
\end{center}
\label{feynman}
\end{figure}

The electron-proton collider HERA has considerably
extended the kinematic
region available to investigations of nucleon constituents.
In particular it permits measurements of processes
directly initiated by gluons in the proton.
In the
boson-gluon fusion process (Fig.~1a) a gluon
from the proton interacts with the virtual boson from the
electron to produce a quark-antiquark pair.
In the range of momentum transfer ($Q^2$) considered in this analysis
photon exchange is dominant and
the contribution from $Z^o$ exchange can be neglected.
Typically the final state
contains two jets in addition to the proton fragment which, to a large
extent, disappears undetected down the beam pipe. Such events are
denoted as (2+1) jet events.
% to account for both the jets of the hard
%subsystem and the spectator jet originating from the proton fragment.
The hard sub-system is the system produced by the interaction of
the virtual photon with a parton in the proton.
The (2+1) jet cross section also receives a contribution from the
QCD-Compton process (Fig.~1b) where a gluon is emitted from the
interacting quark. The cross section of (2+1) jet events
can in leading order schematically be written as
$$
  \sigma_{2+1} \propto \alpha_s (A\cdot g + B\cdot q)
$$
where $\alpha_s$ is the strong coupling constant and
$g$ and $q$ stand for the gluon and quark densities.
The coefficients $A$ and $B$ can be calculated in
pertubative QCD.
The (2+1) jet rate at HERA offers a possibility
to determine
%for the first time in one and the same experiment
the variation of the strong coupling constant with the momentum
transfer of the exchanged boson.
Such a study has been carried out \cite{h1alphas} based on
the assumption that the quark and gluon densities in the
proton are sufficiently constrained at lower momentum
transfer by previous experiments.
The analysis of the gluon density presented here,
on the other hand,
relies upon a reasonably good knowledge about the strong
coupling constant and the quark densities.
%On the other hand a measurement of the
%gluon density from the boson-gluon fusion process relies upon an
%assumption about the value of the strong coupling constant and
%that the quark density is reasonably well known since the quark
%initiated QCD-Compton process constitutes a background source to
%this measurement.
%
%Such a determination has been carried out in a
%kinematic region where the
%quark and gluon initiated processes give equal contributions
%\cite{h1alphas}.
%On the other hand, the analysis described here is performed
%in a kinematic region
%dominated by the gluon initiated boson gluon
%fusion process
%($x$ and $Q^2$ lower than for the $\alpha_s$ study).
%
%The gluon density in the proton is extracted from a measurement
%of the (2+1) jet cross section, assuming that
%the strong coupling constant and the quark
%densities are reasonably well known.
In the kinematic region used for this
analysis,
the multi-jet processes are initiated by partons
carrying a momentum fraction of the proton of the order of
$10^{-2}$, a region in which the quark densities have been accurately
measured in previous experiments \cite{nmcf2,bcdmsf2}.

%The contribution from
%resolved photon processes, where the photon fluctuates into a hadron,
%can be neglected because of the $Q^2$
%requirement in the selection of
%deep inelastic scattering (DIS) events.

\section{The H1 Detector}

A detailed description of the H1 detector can be found elsewhere
\cite{h1}.
%The concept behind the construction design is similar to that of
%other general-purpose detectors at high energy colliders.
Closest to the interaction point there are
tracking devices surrounded by a calorimeter consisting of an
electromagnetic and a hadronic section. Outside these detectors a
superconducting coil provides a magnetic field parallel to the beam
line and, finally, the instrumented magnet iron gives a rough measurement
of the energy leaking out of the calorimeter, and signals the presence
of a muon track.
Here we only give a brief description of
the detector parts which are of relevance to the measurement of the
gluon density.

The backward electromagnetic calorimeter (BEMC) covers the
angular range $155^o<\theta<176^o$ where $\theta$ is defined
with respect to the proton beam direction, which is in the
following called the forward direction.  Making use of the
track information from the backward proportional wire chamber
(BPC), located right in front of the BEMC, and the
reconstructed vertex position, the energy and scattering angle
of the scattered electron can be measured in the kinematic
region considered here.  The BEMC consists of a 22.5 radiation
length deep lead-scintillator sandwich stack, each layer read
out by two pairs of oppositely positioned wavelength shifter
bars. This system gives an energy resolution of
$\sigma (E)/E \approx 0.1/\sqrt{E [\mbox{GeV}]}
\oplus 0.42/E [\mbox{GeV}]
\oplus 0.03$.
By adjusting the measured electron energy
spectrum to the kinematic peak,
the BEMC energy calibration is known to
an accuracy of 1.7$\%$ \cite{f2}.
The BPC has four wire planes, giving a spatial
resolution of 1.5 mm, which together with the precision in the vertex
reconstruction, results in a resolution of the polar angle
of 2.5 mrad.

For the measurement of the hadronic final state
the liquid argon calorimeter was used, which extends over the
angular range $4^o<\theta<153^o$ with a complete azimuthal
coverage. The total depth varies between 4.5 and 8 interaction
lengths. Measurements on test beams have given a hadronic energy
resolution of $\sigma (E)/E \approx 0.5/\sqrt{E [\mbox{GeV}]} \oplus
0.02$ \cite{hadscale}. The absolute hadronic energy scale has
been measured to a precision of 5$\%$ from the transverse
momentum balance between
hadronic jets and the scattered electron in deep-inelastic scattering
(DIS) events.

\section{Kinematics}

In 1993 HERA was operated to collide 26.7 GeV electrons
with 820 GeV protons, resulting in a center-of-mass
energy of 296 GeV.  $\:$Fig.~1 shows Feynman diagrams of
the boson-gluon fusion process (BGF) and the QCD-Compton
process (QCD-C) with the relevant kinematic variables
indicated.

The overall kinematics of an event can be
determined from two independent Lorentz invariant variables.
This can be any two of
the Bj$\o$rken scaling variables $x$ and $y$, the photon
momentum transfer squared $Q^2$, and the
invariant mass squared of the hadronic system $W^2$.
These variables are defined in terms of
the four-momenta of the incoming proton, $P$, the incoming and
outgoing electron, $p_e$ and $p'_e$, and the exchanged photon, $q$.
Experimentally they are deduced from measurements of the energy,
$E'_e$, and polar angle, $\theta_e$, of the scattered electron
according to the following
relations (neglecting the electron and proton masses):

%\begin{equation}
%\begin{array}{l}
\[
Q^2 \equiv -q^2 = -(p_e-p'_e)^2 = 4E_eE'_e\cos^2(\theta_e/2)
\]
\[
y \equiv \frac{P \cdot q}{P \cdot p_e} =
1 - (E'_e/E_e)\sin^2(\theta_e/2)\]
\[
x \equiv \frac{-q^2}{2P \cdot q} = \frac{Q^2}{ys}; \hspace{1cm}
W^2=Q^2\left(\frac{1-x}{x}\right)
\]
%\end{array}
%\end{equation}
%
The center-of-mass energy squared is given by $s = 4E_eE_p$,
where $E_e$ and $E_p$ are the energies of the incoming electron
and proton, respectively.  Note that Bj$\o$rken-$x$ is not
identical to the fractional momentum $x_{i/p}$ of the parton
$i$ initiating the hard sub-process, with $i$ standing for
either a quark $(q)$ or a gluon $(g)$.  $x_{i/p}$ is related to
the invariant mass squared of the hard sub-system ($\hat{s}$) in
the following way:
\begin{equation}
x_{i/p} = \frac{\hat{s}+Q^2}{ys} = x\left(1+\frac{\hat{s}}{Q^2}\right)
\label{eq:xip}
\end{equation}
where $\hat{s}$ can be measured either directly from
the invariant mass
of all particles $p_j$ of the two jets belonging to
the hard sub-system
\begin{equation}
%\begin{array}{l}
\hat{s} = \left( \sum_{j} p_j \right)^2
\label{eq:energy}
%\end{array}
\end{equation}
or from the jet directions
in the
hadronic center-of-mass
system, which is the rest system of the exchanged photon and the
proton:
\begin{equation}
%\begin{array}{l}
\hat{s} =
W^2 e^{-(\eta_{1}^*+\eta_{2}^*)}
\label{eq:rapidity}
%\end{array}
\end{equation}
where $\eta_{1}^*$ and $\eta_{2}^*$ are the parton pseudo-rapidities
of the hard sub-process\footnote{
Variables $v$ in the hadronic center of mass system are denoted as
$v^*$ while variables in the photon-parton
center of mass frame are denoted as $\hat{v}$.
The direction of the  $z$-axis in the hadronic center of mass system
is defined as the direction of the exchanged photon.
Note that the transverse momentum $p_T^* = \hat{p}_T$.}
and $\eta= - \ln\tan(\theta/2)$.
Experimentally the parton directions and momenta are approximated
by the directions and energies of the jets produced in the
fragmentation process.

\section{Trigger and Data Selection}

The analysis is performed with deep-inelastic scattering (DIS) events
which are selected in the detector by
applying the trigger requirement for an electron of a local energy
deposition (cluster)
of more than 4 GeV in the BEMC detector.
%Fake DIS events resulting from the proton
%beam induced background upstream of the detector are
%removed already at the trigger level by using flight time measurements
%from two planes of  scintillators       behind the BEMC.
The total event sample used in this analysis
corresponds to an integrated luminosity of 242 nb$^{-1}$ from the
data sample collected in 1993.
After the event reconstruction has been performed,
a clean DIS sample is obtained by applying the following
requirements:

$\bullet$ The scattered electron energy had to satisfy $E'_e>10
\mbox{GeV}$ which corresponds to $y \lsim 0.625$. This removes
most of the background due to the photoproduction reactions
where an electromagnetic shower in the hadronic final state is
misidentified as an electron from the DIS process.
%The events
%of this type contributing to our data sample have the electron
%faked by the hadronic final state.

$\bullet$ The track coordinates in the BPC had to match the cluster
center-of-gravity
in the BEMC to within 4 cm and the lateral spread of the cluster had to
be less than 4 cm in radius.

$\bullet$ The momentum transfer squared had to be in the range
$12.5<Q^2<80 \mbox{ GeV}^2$ which ensures that the scattered
electron is well contained in the BEMC.

$\bullet$ In order to get an adequate resolution in $y$ from the
measurement of
the scattered electron it was required that $y > 0.05$.

$\bullet$ At least one charged track from the hadronic final state
was demanded for the determination of
the vertex position along the beam,
which had to be within 30 cm from the
nominal position in order to suppress beam-induced background.

After these cuts and the jet selection described below
no indication was observed of remaining background from
photoproduction in our data. This was confirmed by a Monte Carlo
simulation of photoproduction processes.

Efficiencies for passing the above cuts have been determined
from data using the redundancy in our apparatus.  Further
details can be found in \cite{f2}. The $4 \%$ error in the
combined efficiency will contribute to the overall
normalization error.

The (2+1) jet selection is carried out in two steps.
In the first step the jet finding is performed
by applying a cone algorithm \cite{cone}
on calorimeter clusters with
$\Delta R= \sqrt{\Delta{\eta^2} + \Delta{\Phi^2}} = 1$
first boosting to the hadronic center-of-mass system.
$\Delta\eta$ and $\Delta\Phi$ are the differences in pseudo-rapidity
and azimuthal angle between two cluster pairs respectively.
Exactly two jets with a transverse jet energy $p_T^*>3.5
\mbox{GeV}$ are required.  The hadronic center-of-mass
system is a natural frame for a QCD analysis as it is not
sensitive to the $p_T$ of the hadronic system due to the
transverse momentum balance of the electron in the
laboratory system.  It is also a preferred frame for the
$p_T$ based cone algorithm. The ratio of (3+1) to (2+1)
jet events is less than 15\%.

In the second step following cuts are imposed on the jet
system of the event:

$\bullet$
Both reconstructed jets have to fall inside the angular range
$10^o<\theta_{jet}<150^o$ in the laboratory system
so as to be within the volume covered by the liquid argon
calorimeter.
%The region above 140$^o$ which provides a poorer
%measurement of the hadronic energy contains less than 1\% of
%the events.
A further purpose of the lower cut in angle is to
remove the very forward region which is dominated by the
proton fragments and initial state parton radiation which
may otherwise give rise to a separate jet.

$\bullet$
The difference in pseudo-rapidity between the two jets
in the laboratory frame must
satisfy $\Delta\eta<2$.
This corresponds to a cut on the scattering angle in the
photon-parton center of mass system of $|\hat{\eta}| \lsim 1$ or
$40^o \lsim \hat{\theta} \lsim 140^o$ in the kinematic range
used for this analysis.
Note that this cut, except for the highest $x_{g/p}$
covered by this analysis,
is more restrictive than the angular cut discussed above.

$\bullet$ The corrected invariant mass (see section 6) of
the two jets of the hard scattering system is required to
be $\sqrt{\hat{s}}>10$ GeV to ensure well defined jet
structures (see Fig.~2).  This cut, together with the jet
angular cut $\Delta\eta<2$, implies $p_T^* \gsim 3.2$ GeV
and is therefore well matched to the $p_T^*$ cut of 3.5
GeV in the application of the cone algorithm.  In addition
both methods are used to reconstruct $\hat{s}$, as given
by equations (2) and (3). It is required that the
difference in $|\Delta \sqrt{\hat{s}}|$ for the two
methods is smaller than $10 \mbox{ GeV}$.

With these cuts a sample
of 328 (2+1) jet events covering the
fractional momentum range $0.002 \lsim x_{g/p} \lsim 0.2$
and the Bj$\o$rken-$x$ range $0.0003 \lsim x \lsim .0015$
is obtained.
%It has contributions from both
%BGF and     QCD-Compton processes.
%Since we have found no efficient method to
%distinguish between BGF and QCD-Compton events we have to restrict
%ourselves to a kinematic region which is dominated by the BGF process.
In the kinematic region which was chosen for this analysis,
according to the cuts defined above,
the ratio of BGF and QCD-Compton
cross sections is approximately 3/1, independent of $x_{i/p}$.

\section{Monte Carlo Generation}

\begin{figure}[htbp]
\vspace*{-2.5cm}
\begin{center}
\begin{tabular}{c}
  \mbox{\epsfig{file=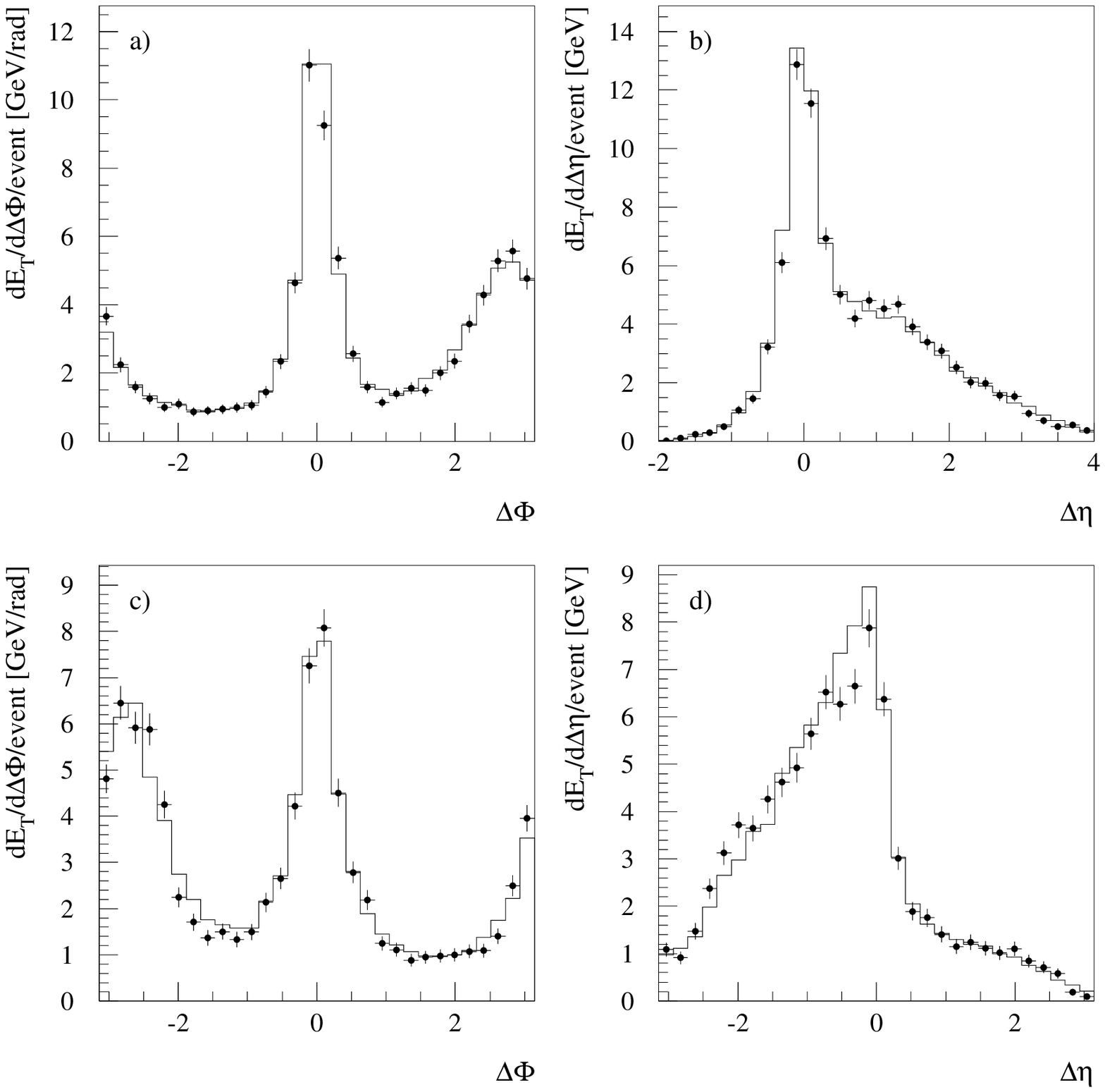,height=15cm}} \\[-1.3cm]
  \mbox{\epsfig{file=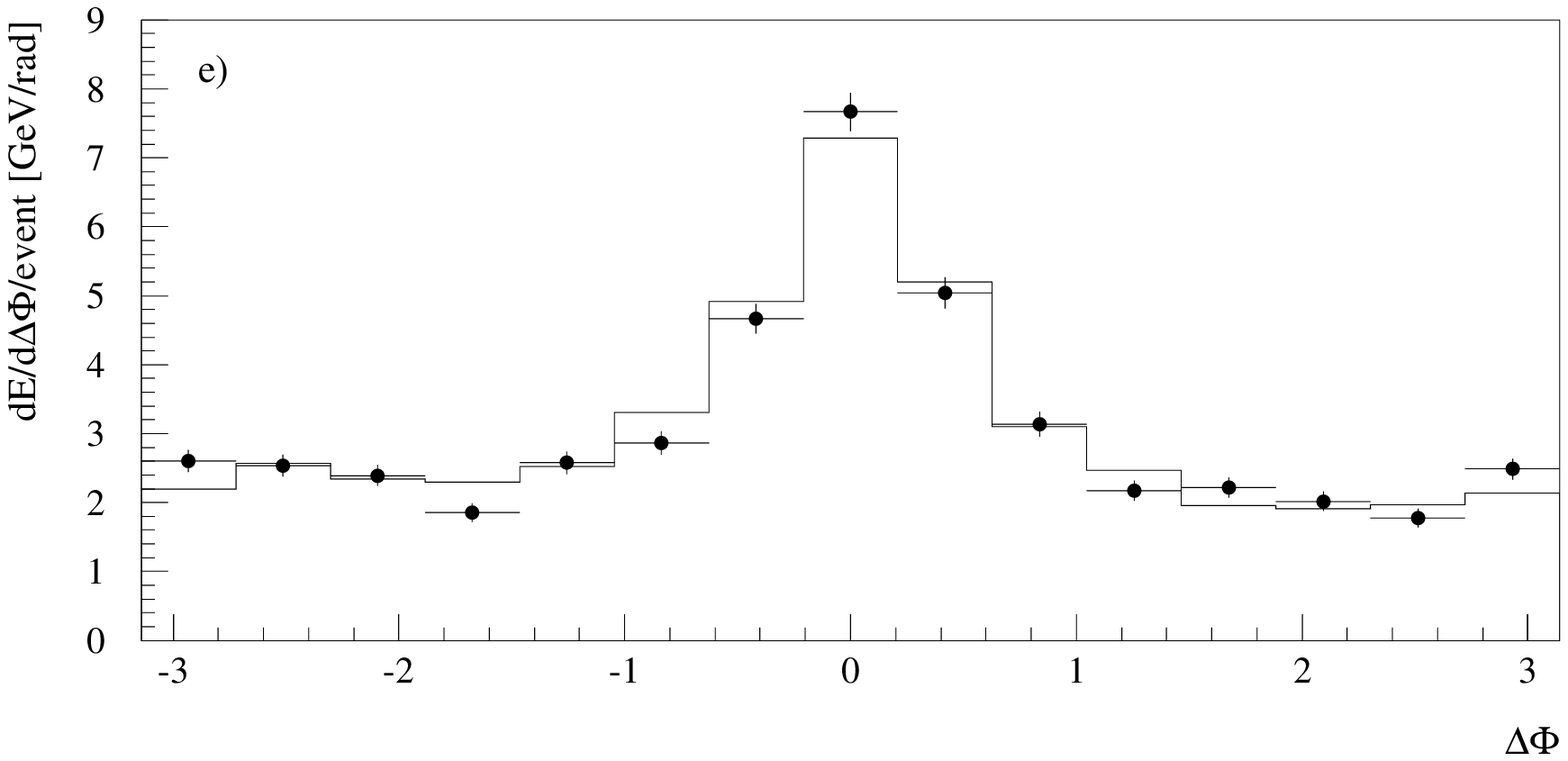,height=15cm}} \\[-8.0cm]
\end{tabular}
\end{center}
\caption[control2]{{\it
The transverse energy flow in the
laboratory frame as a function of a) $\Delta \Phi$ and
b)  $\Delta \eta$ for the most backward going jet, and
as a function of c) $\Delta \Phi$ and
d) $\Delta \eta$ for the most forward going jet, using the
respective jet axis as reference.
The energy flow in the region $2< \eta <3$ is shown with
the jet axis of the most forward going jet as reference in e).
The sense of $\Phi$ is away from the scattered electron for
the jet at the smallest $\eta$.
The points represent the
(2+1) jet data sample and the errors are statistical whereas
the histograms give the predictions of the MEPS model.}}
\label{control2}
\end{figure}

In order to calculate cross sections from matrix elements
and to correct for parton shower, hadronisation, and
detector effects, the Monte Carlo event generator LEPTO
6.1 \cite{lepto} was used, with the parton density
function MRS H \cite{MRS}.  Although it would have been
more natural for this analysis to use leading order
parametrisations of the density functions it will be
explained in the discussion of errors (section~7) that
this is not essential.  The strong coupling constant
$\alpha_s$ was calculated in first order QCD with
$\Lambda_{QCD}$ = 250 MeV for 5 flavours.  It was checked
that LEPTO delivers cross sections which are consistent
with leading order cross section predictions by the
PROJET \cite{projet} and DISJET \cite{disjet} programs
within 5\% in the kinematic region covered by this
analysis.

The LEPTO generator is based on QCD matrix element
calculations up to first order in $\alpha_s$, and the inclusion
of parton showers accounts for higher order processes (MEPS).
The QCD
matrix element gives divergences for soft and collinear emission
which is technically avoided by defining a smallest invariant mass
$m_{ij}$ between any two partons, including the remnant.
The parametrisation used for the cut in $m_{ij}$
is always 2 GeV above the region in phase space where
the 2+1 jet cross section would exceed the total cross section.
%We have used a parametrisation for the cut in $m_{ij}$
%which follows the divergency limit and is made such that
%$m_{ij}$ exceeds it always by 2 GeV.
%%We have used a parametrisation for the cut in $m_{ij}$
%%which follows with a 2 GeV margin above
%%the divergence limit.
%At the divergence limit,
%the 2+1 jet cross section exceeds the total cross section.
With such a cut, matrix element calculations are used to generate
events over the maximum possible phase space, thereby giving access
to regions of small $x_{g/p}$.
The Monte Carlo matrix element cut-off is always
significantly below 10 GeV which is the constant mass cut
applied in the event selection.
Therefore the full phase space used in this analysis is
covered by the matrix element.
There is, however, some migration of events with invariant
masses below 10 GeV. The contribution from the zeroth order
process (QPM events) is about 60\% at the generator level.
In the final event sample a 2\% contribution from
these processes is observed, generated as (1+1) jet
events, but wrongly reconstructed as (2+1) jet events.
This indicates that the contribution from the phase space
not covered by the matrix element is small.

Although the matrix element calculation used in LEPTO
assumes massless quarks, the program contains a
parametrisation which takes the heavy quark mass thresholds
into account.  In the kinematic region of this analysis,
this effect reduces the LEPTO BGF cross section by 11\% on
the average.

A generated sample corresponding to about four times the
statistics of the data was subject to full detector simulation,
followed by event reconstruction.

%In order to gain confidence in the results provided by the Monte Carlo
%generator
A number of control plots have been made to compare
the predictions of the Monte
Carlo program with the
experimental data.
Fig.~\ref{control2}a-d show the transverse energy flow as a function
of the azimuthal angle, $\Phi$, and pseudo-rapidity, $\eta$, for the
two jets using the respective jet axis as reference.
%The well described transverse energy flow in the backward
%region ($\Delta\eta < 0$ in Fig.~\ref{control2}b)
%indicates that there is
%no significant contribution from resolved processes as expected
%due to the $Q^2$ requirement in the selection of the DIS sample.
Particular attention was paid to
the Monte Carlo
description of the transverse energy flow in the forward region,
$2< \eta <3$, which might contain a large contribution
from initial state QCD radiation.
Although we have previously observed
\cite{eflow} that the MEPS model is not able to reproduce the
energy flow in this rapidity region for an inclusive DIS sample,
it is demonstrated
in Fig.~\ref{control2}e that the energy flow of the selected
(2+1) jet sample
is well described by the MEPS model. It can be noted that not only the
jet profile exhibits good agreement between data and the
Monte Carlo sample but also the level of the underlying energy flow.
Other control plots, such as the rapidity distributions of the two
hard jets and their transverse energy spectra, as well as the
$Q^2$, $W^2$, $y$ and $x$ distributions,
give further evidence for the ability of the MEPS model
to reproduce the data.

In order to estimate the model dependence of the BGF
acceptance LEPTO MEPS was compared with other models
available. It was found that the MEPS-, CDM 4.03- (colour
dipole model) \cite{cdm} and HERWIG 5.7 \cite{herwig} models
all give acceptances for BGF-events agreeing within 15\%
at the hadron level.
%In addition, HERWIG provides an acceptance for the
%QCD-Compton background which agrees with the MEPS number to
%within 15\%.
The CDM model also gives a good description of
the jet profiles but it is not able to give a satisfactory
description of the cross section as a function of kinematic
variables and jet variables as previously observed with less
statistics in \cite{H1CDM}.
The discrepancies originate from non-BGF processes.
In HERWIG there is no matching between the zeroth and first
order processes, preventing an estimate of the migration
of background.

%In addition, HERWIG also provides an acceptance for the QCD-Compton
%background. It also agrees with the MEPS result within 15\%.

\section{Reconstruction of $x_{g/p}$
and Unfolding of the Gluon Density}
\label{xgprec}

\begin{figure}[t]
\vspace*{-1.5cm}
\begin{center}
\mbox{
\epsfig{file=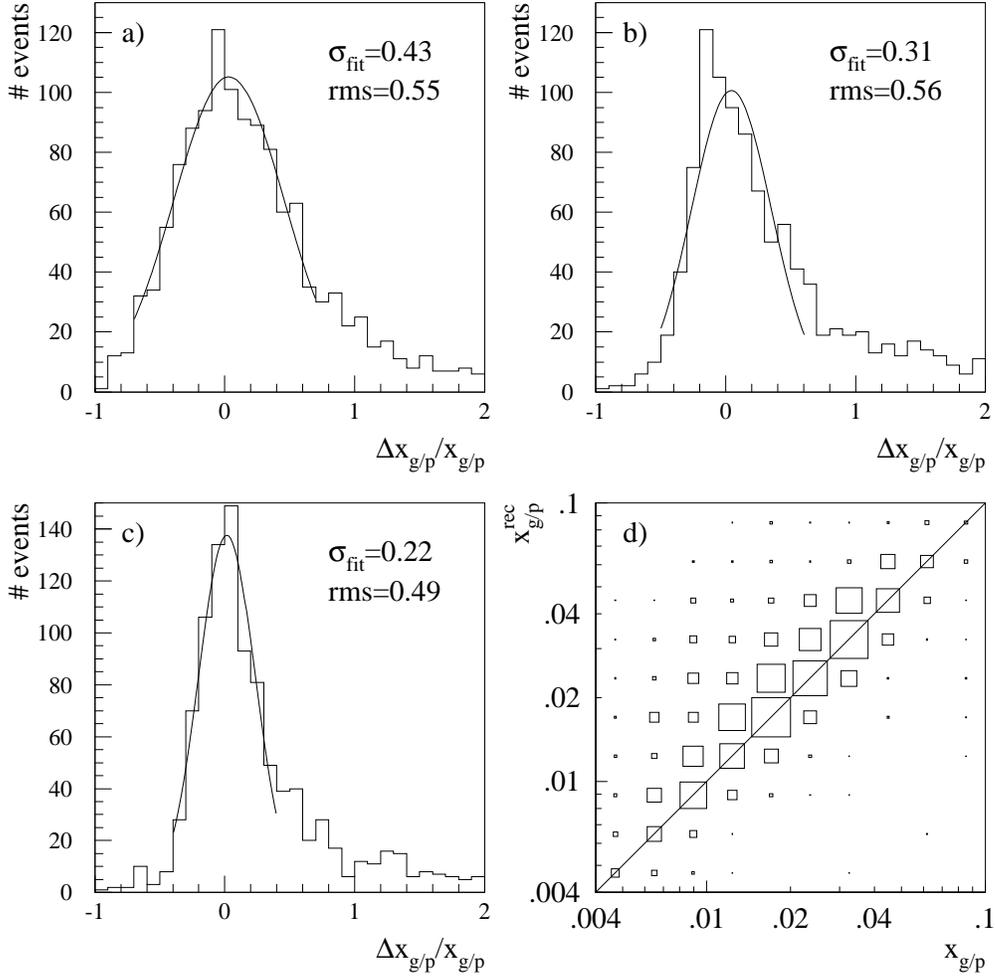,height=14.8cm}}
\end{center}
\vspace*{-1cm}
\caption[xgrec]{{\it
The relative error in the reconstruction of the fractional gluon
momentum, $\Delta x_{g/p}/x_{g/p}$, where $\Delta x_{g/p} =
x_{g/p}^{rec} - x_{g/p}$, for BGF events generated by the
MEPS model. In a) equation (\ref{eq:energy}) has been used,
in b) equation (\ref{eq:rapidity})
and in c) the combination of the two.
%The reconstruction of $x_{g/p}$ using a) equation (\ref{eq:energy}),
%b) equation (\ref{eq:rapidity}) and c) the combined method.
The correlation between the reconstructed and true $x_{g/p}$ for
the combined method is shown in d).}}
\label{xgrec}
\end{figure}
The Monte Carlo sample was used to investigate the
correlation between $x_{g/p}$ of the gluon calculated from the
hard partons originating from the QCD matrix element
%as obtained by the
%the $x_{g/p}$ of the hard partons
%originating from
%QCD matrix element
and $x_{g/p}^{rec}$
calculated from the jets measured in the detector.  It is
essential to use a jet reconstruction algorithm which
provides good separation between the spectator jet and the
jets of the hard sub-system. The event sample populates the
region of small $x$ and $Q^2$ values, the spectator jet will
carry away most of the available energy.  A misassignment of
one of the relatively energetic particles from the spectator
jet into the hard sub-system has a strong impact on the
reconstructed $\hat{s}$ value and will distort the $x_{g/p}$
reconstruction. It is therefore important to optimize the
resolution parameters of the jet algorithms to give the best
possible performance in this respect.

The jet definition and detector effects lead to systematic
shifts of the order of 30\% in the determination of
$x_{g/p}$ and $\hat{s}$ (see equation \ref{eq:xip}).  The
shifts have been determined separately for the two methods
of extracting $x_{g/p}$ (equations (\ref{eq:energy}) and
(\ref{eq:rapidity})).  Figs~\ref{xgrec}a and b show the
relative error in the $x_{g/p}$ reconstruction after correcting
for the shift.

Comparing the results of the two methods it can be seen
that the fitted Gaussian distributions result in a
superior resolution for the method based on equation
(\ref{eq:rapidity}) which, however, exhibits a more
pronounced tail.

For properly reconstructed events both methods are expected
to give consistent results, whereas
a misassignment of particles by the jet algorithm might have
different impacts on the $\hat{s}$ (and therefore
$x_{g/p}$) reconstruction from the two
methods.
Also for (1+1) jet events, where an additional jet arises from
high $p_T$ initial parton emission, the methods
do not give the same values of $\hat{s}$.
%do not give a consistent measurement of $\hat{s}$.
%
Therefore an improved result is        obtained by selecting only
events for which the result of the two methods agree within
the resolution.
It has been demanded that the
absolute value of
the difference between
the $\sqrt{\hat{s}}$ values extracted from equations
(\ref{eq:energy}) and (\ref{eq:rapidity}) is
$|\Delta \sqrt{\hat{s}}| \leq 10 \mbox{ GeV}$ and then simply
taken the mean
value of the two reconstructed values to give the combined
$x_{g/p}$. Due to this cut about 20\% of the events are removed.

As shown in Fig.~\ref{xgrec}c, the result of the combined method
shows a considerably improved resolution compared with the methods
based on equations (\ref{eq:energy}) and (\ref{eq:rapidity}) separately.
In Fig.~\ref{xgrec}d the correlation between the reconstructed
and true $x_{g/p}$ is given for the combined method.
{}From here on the combined method is used to extract $x_{g/p}$.
%Subsequently, we will use the combined method to extract $x_{g/p}$.

%section{Unfolding of the Gluon Density
%g(x_{g/p},Q^2)$}

%The acceptance defined as
%the ratio of the number of
%BGF events accepted after detector simulation and
%reconstruction and the number of BGF events generated in the
%acceptance region (equations \ref{eq:accreg}) is 40\%.

\begin{table}[b]\begin{center}
\begin{tabular}{|c|c|c|c|c|c|}
 \hline
 & & & & diffractive \\
  $x_{g/p}$-bin range &
$ x_{g/p}g(x_{g/p})$ &
$<Q^2>$ [GeV$^2$] &
$<p_T^{*}>$ [GeV] &
contribution [\%] \\
 \hline
  0.0019 - 0.0061 &
  $9.04 \pm 2.95$ &
  23 &
  5.0 &
  $16 \pm 9$ \\
  0.0061 - 0.012 &
  $7.40 \pm 1.28$ &
       29 &
        5.4 &
   $10 \pm 5$ \\
  0.012 - 0.030 &
  $4.03 \pm 1.06$ &
            30 &
              6.0 &
   $6 \pm 3$ \\
  0.030 - 0.052 &
  $1.42 \pm 0.60$ &
                 35 &
                    6.6   &
   $9 \pm 4$ \\
  0.052 - 0.18  &
  $1.18 \pm 0.80$ &
                      36 &
                            9.8   &
   $9 \pm 6$ \\
\hline
\end{tabular}
\end{center}
\caption[]{\it Results and kinematic characteristics. Errors
include statistical and systematic uncertainties, excluding a
11\% global  normalization uncertainty.}
\label{tabres}
\end{table}

% NEW UNFOLDING PART !
The observed (2+1) jet cross section as a function of
$x_{g/p}^{rec}$ is interpreted as the sum of a gluon
initiated (BGF) and a quark initiated (QCD-C plus QPM)
part, according to
\[
\sigma^{2+1}_{obs.}(x_{i/p}^{rec}) =
\int M(x_{g/p},x_{g/p}^{rec})
x_{g/p} g(x_{g/p}) \mbox{d}x_{g/p}~~~+
\sigma^{QCD-C}_{MC}(x_{q/p}^{rec}) +
\sigma^{QPM}_{MC}(x_{q/p}^{rec})
\]
The quark initiated contributions represent the
background obtained from the LEPTO Monte Carlo program.
The integral represents the BGF part and expresses the
convolution of the gluon density $x_{g/p} g(x_{g/p})$
with the function $M$ which besides the measurement
process describes effects due to the QCD matrix element,
parton showering and hadronisation. The unfolding
procedure with regularisation described in \cite{blobel}
is used to determine the gluon density.  The $x_{g/p}$
range covered by the kinematic region is subdivided into
five bins.  The bin boundaries (see Tab.~1) were
optimized in order to minimize the correlations between
the bins.

\begin{figure}[t]
\vspace*{-1.0cm}
\begin{center}
\mbox{\epsfig{file=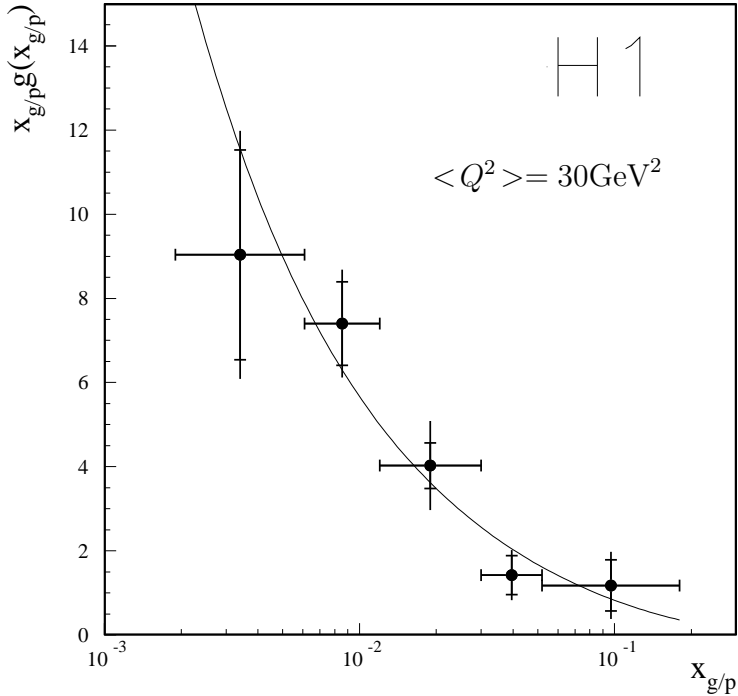,height=10.5cm}}
\put(-130,200){$<\!Q^2\!>=30\mbox{GeV}^2$}
\caption[gdens]{{\it
The measured gluon density at an average $Q^2$
of 30 GeV$^2$ as a function of the fractional gluon
momentum. The error bars reflect the
statistical errors and the total errors respectively.
Not included is the global normalization uncertainty of
11\%. The solid curve shows a fit to the data points
as explained in the text.}}
\end{center}
\end{figure}

The results for the gluon density as a function of the
fractional gluon momentum are presented in Fig.~4.  The
inner and outer error bars on the data points reflect the
diagonal elements of the statistical and full covariance
matrix respectively.  Not shown is the 11\% global
normalization uncertainty.  Additional information is
given in Table 1.  Results are quoted here at a scale of
30 GeV$^2$ corresponding to the average $Q^2$ value of
the data sample.  Another possible choice for the
QCD-scale would be the average $p_T^{*2}$ of the jets in
the hadronic center-of-mass system, which for the present
data sample is 40 GeV$^2$. Comparing the GRV gluon
parametrisation \cite{GRV} (see below) for $Q^2$ values of 20 and 40
GeV$^2$ gives less than 10\% variation in our range of
$x_{g/p}$.

A considerable rise in the gluon content of the proton
$x_{g/p}g(x_{g/p})$ with decreasing fractional momentum of
the gluon is observed.  In Fig.~4 a fit of the data points
to the parametrisation $x_{g/p}g(x_{g/p}) =A_g
x_{g/p}^{B_{g}} (1-x_{g/p})^{C_{g}}$ is shown taking the
full covariance matrix into account (see Tab.~2).  The
parameter $C_g$ was kept fixed at the value of 5. The
values of the other parameters as obtained from the fit are
$A_g=0.33\pm 0.16$ and $B_g=-0.63\pm 0.12$ with
$\chi^2/n_f=0.99$. A value of $C_g=9$ lead to variations
of $A_g$ and $B_g$ within the stated errors.

\section{Discussion of Errors}

The unfolding yields a full covariance matrix
of statistical errors which includes information on correlations
between bins due to migration effects.
In this analysis, the correlations are small (see Tab.~2),
indicating adequate resolution
for choosing five bins in $x_{g/p}$.
\begin{table}[b]\begin{center}
\[
\begin{footnotesize}
\hspace*{-4mm}\left( \begin{array}{rrrrr}
     6.220 &       &       &       &       \\
     0.087 & 0.978 &       &       &       \\
    -0.257 &-0.064 & 0.293 &       &       \\
     0.033 &-0.065 &-0.017 & 0.214 &       \\
     0.032 & 0.013 &-0.049 & 0.003 & 0.366
\end{array} \right)
\hspace*{5mm}
\left( \begin{array}{rrrrr}
     8.670 &       &       &       &       \\
     0.858 & 1.632 &       &       &       \\
     0.976 & 0.584 & 1.117 &       &       \\
     0.243 & 0.038 & 0.050 & 0.354 &       \\
    -0.407 & 0.108 &-0.162 & 0.044 & 0.642
\end{array} \right)
\end{footnotesize}
\]
\end{center}
\caption[]{\it Covariance matrix of statistical errors (left) and total
(sum of statistical and systematic)
errors (right). Not included is a global
normalization uncertainty of 11\%.}
\label{covar}
\end{table}
For a quantitative estimate of systematic uncertainties
we varied several parameters and repeated the full analysis
including the unfolding step.
For each variation, a covariance matrix was calculated.
The total systematic covariance matrix
is the sum of the matrices of the individual variations.
The following effects were studied.

The absolute hadronic energy calibration
(affecting the measurement of $p_T^*$ and $\hat{s}$)
was varied by its present
precision of
5$\%$.
We also varied the
electron energy calibration
(affecting the measurement of the Bj$\o$rken variable $y$ and
the total hadronic invariant mass $W$)
by
its measured precision of
$\pm 1.7 \%$.

The jet transverse momentum cut
in the cone algorithm used for jet finding was varied
between 3 and 4 GeV.
The jet angular cut
$\Delta \eta$ nominally 2 was varied between 1.5 and 2.5.

In section 5 it was shown that the MEPS model gives an
excellent description of the data sample used in this
analysis.  However, comparing the distribution of the
absolute difference of $\sqrt{\hat{s}}$ reconstructed
according to equations (2) and (3) ($|\Delta
\sqrt{\hat{s}}|$, see section 6) an excess of data events
is observed compared to the MEPS model at values of
$|\Delta \sqrt{\hat{s}}|$ larger than 15 GeV.  Below this
value Monte Carlo and data show good agreement.  The
excess in data is due to events with at least one jet in
the forward region, a region where the Monte Carlo
description of data has been shown to be problematic
\cite{eflow}.  However, even for the events rejected by
the cut $|\Delta \sqrt{\hat{s}}| < 10$ GeV the MEPS model
gives a good description of the energy flows.  For
systematic studies, the $|\Delta
\sqrt{\hat{s}}|$ cut is varied between 7 and 15 GeV.

QCD-Compton cross sections using two leading order (LO)
quark density para\-me\-tri\-sations (GRV \cite{GRV} and
CTEQ3L \cite{CTEQ})
have been compared with the next to leading order (NLO)
MRS H function used in the MC model.  The discrepancies
are largest at the low $x_{q/p}$ but do not exceed 15\%
in the kinematic domain used.  As a conservative estimate
of the systematic uncertainty related to the subtraction
of the QCD-Compton contribution a variation of $\pm 25\%$
was used to determine the contribution to the systematic
error.

Tab.~2 shows the total covariance matrix which is the sum of
the statistical and systematic covariances.  The relative
systematic errors vary between 17\% in the lowest bin of
$x_{g/p}$ and 44\% in the highest bin.  The dominating
contribution to the systematic errors in the lowest $x_{q/p}$
bin is 13\% coming from the uncertainty in the hadronic energy
calibration of the liquid argon calorimeter, while the
dominating systematic error in the highest $x_{q/p}$ bin is
35\% due to the variation of the $|\Delta \sqrt{\hat{s}}|$ cut.

In addition there is a global normalization uncertainty of 11\%
which arises from the following sources: a 4.5\% uncertainty in
the luminosity measurement \cite{lumi} directly propagates into
the gluon density measurement, as does a 4\% uncertainty in
various detector efficiencies mentioned in section 4.  Finally,
the one standard deviation uncertainty in the strong coupling
constant $\alpha_s$ at $Q^2 = M^2_Z$ \cite{alphas} contributes
a 9\% uncertainty at $Q^2 = 30$ GeV$^2$.

The variation of the parameters of the unfolding procedure
within wide ranges had a negligible effect on the result.

The Monte Carlo model dependence of the BGF and QCD-Compton
acceptances was estimated on the hadron level by comparing the
numbers obtained from the CDM, HERWIG and MEPS models and was
found to be below 15\% (see section 5).

The matrix element cut-off in the Monte Carlo generator
on the hadron level was varied by $\pm$1 GeV in order to
study how it affects the event composition. No
significant change was observed.

The radiative correction has been estimated for each
$x_{g/p}$ bin by calculating the cross section with and
without including QED radiation in the DJANGO\,2.1 Monte
Carlo program \cite{django}.  This led to variations
consistent with the statistical accuracy of the generated
Monte Carlo samples.

The full analysis was repeated using the JADE jet finding
algorithm \cite{JADE} in the laboratory frame with a
fixed mass cut of 10 GeV. This is an algorithm based on a
fundamentally dif\-fe\-rent principle of reconstructing
jets compared to the cone algorithm which results in
different numbers of selected events and different
migrations. In spite of this the finally obtained results
on the gluon density agree within the statistical
accuracy.  The reason for using the cone algorithm to
extract the final results was the better $x_{g/p}$
resolution and the greater suppression of tails.

As a further systematic check the analysis was repeated
with a stricter cut in the invariant mass of the hard
sub-system $(\hat{s} > 200 \mbox{GeV}^2)$, again yielding
variations well within the statistical accuracy,
although the statistical errors were significantly increased.

\section{Discussion of the Results}

\begin{figure}[t]
\vspace*{-1.0cm}
\begin{center}
\mbox{\epsfig{file=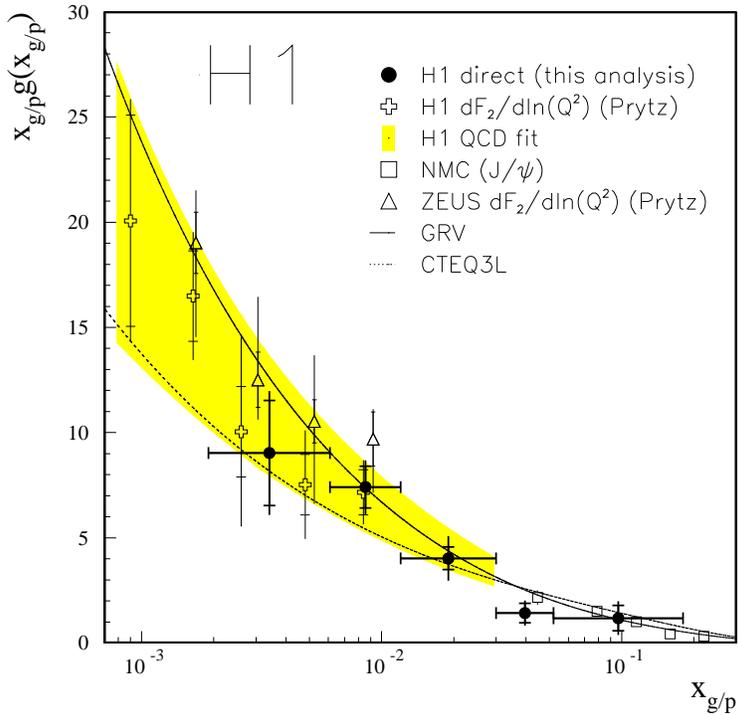,height=10.5cm}}
\caption[gdens]{{\it
The measured gluon density at an average $Q^2$ of 30 GeV$^2$
as a function of the fractional gluon momentum compared with
indirect determinations by H1 \cite{H1new} and ZEUS
\cite{ZEUS} at $Q^2$ = 20 GeV$^2$ as well as with a
determination from $J/\Psi$ production
by NMC \cite{NMCdir} evolved to
$Q^2$ = 30 GeV$^2$ (see
text).}}
\end{center}
\end{figure}

So far only indirect constraints on the gluon density
based on a $Q^2$ evolution of the quark density have been
obtained for low values of $x$.  A comparison between such
indirect extraction and direct measurement of the gluon
density constitutes an important test of perturbative QCD.
Fig.~5 shows a comparison of this measurement with recent
indirect determinations by the H1 \cite{H1new} and ZEUS
\cite{ZEUS} collaborations at a $Q^2$ of 20 GeV$^2$ which
covers an even lower $x_{g/p}$ region. Our results are
also consistent with a previous indirect measurement by
NMC \cite{NMCin} in a region $x_{g/p}>10^{-2}$.

Direct measurements have been reported by UA2 \cite{UA2}
and NMC \cite{NMCdir} at $x_{g/p}$ values above 0.04.  The
NMC data based on inelastic J/$\psi$ production in DIS is
included in Fig~5.  The parametrisation of the NMC data
provided by this collaboration was used to perform a LO
DGLAP evolution from $Q^2 = 1.5$ GeV$^2$ to $Q^2 = 30$
GeV$^2$.

Further the results are compared with two different
parametrisations of the gluon density in LO.  The GRV
 model assumes the gluons and sea quarks to be
valence-like at $Q^2$ = 0.23 GeV$^2$ and the growth of the
gluon density with decreasing $x_{g/p}$ values is due to
radiation of low $x$ partons generated according to the
Altarelli-Parisi evolution equations.  The CTEQ3L
 parametrisations are based on input
distribution functions at $Q^2 = 4$ GeV$^2$ assuming the
sea and gluon distributions to have the same power
dependence of the fractional momentum
($x^{B_{g}}$).  Both are consistent with
our result.

The contribution of diffractive events, in which the
exchanged boson interacts with a colourless object in the
proton and therefore does not span a colour string between
the hard sub-system and the proton, has been evaluated
using data by selecting events with no activity in the
forward detectors \cite{diffr}.  After correcting for the
efficiency of the diffractive selection and for non
diffractive events passing the selection cuts, a $(8 \pm
2)$\% contribution is observed in the data sample
consistent with a flat $x_{g/p}$ dependence (see also
Table 1).

Next to leading order (NLO) corrections using the cone
algorithm for jet production in deep-inelastic scattering
do not exist.  In photoproduction these corrections have
been found to be of the order 25\% for an average
$<p_T^*>$ (see Tab.~1) similar to this analysis \cite{gp}.
Although in photoproduction besides the contribution from
BGF and QCDC the so called resolved processes have been
taken into account, NLO corrections of similar size are
expected \cite{sg}.

\section{Conclusions}

{}From a measurement of the cross section
for (2+1) jet events,
a direct LO determination of the gluon density in the proton
has been performed
%
%we have made a direct LO determination of the gluon
%density in the proton
in a previously inaccessible
domain of the gluon fractional momentum $x_{g/p}$
($1.9 \cdot 10^{-3} < x_{g/p} < 0.18$) at an average
value of $Q^2=30$ GeV$^2$.
Our data are consistent with a steep
rise in the gluon density as
$x_{g/p}$ decreases.
%We have estimated the size of the NLO correction to be of the order
%20\% - 30\%.
%We observe a 9\% diffractive
%contribution in our data sample, independent of
%$x_{g/p}$.
Recent indirect extractions of the gluon density
based on the $Q^2$ evolution of the structure
function $F_2$
by the ZEUS and H1 experiments are compatible with our data.

%A comparison has been made between our data and two
%existing parametrisations of the gluon density in leading order.
%The CTEQ3L parametrisation is disfavoured by our data points,
%whereas a more precise conclusion about the GRV model
%will require more statistics and, above all, a better understanding
%of the systematic errors.

\vspace{5mm}
{\bf Acknowledgments:}
We are grateful to the HERA machine group whose outstanding efforts
made this experiment possible. We appreciate the immense effort
of the engineers and technicians who constructed and maintained
the detector. We thank the funding agencies for financial support.
We acknowledge the support of the DESY technical staff. We also wish
to thank the DESY directorate for the hospitality extended to the
non-DESY members of the collaboration.
We have profited from many constructive
discussions with members of the Uppsala and Lund theory groups.

\end{document}